\documentclass[10pt, final]{IEEEtran}

\usepackage{amsthm}

\usepackage{amssymb}
\usepackage{amsmath}
\usepackage{bbm}
\usepackage[ruled]{algorithm2e}
\usepackage{mathrsfs}
\usepackage{cite}
\usepackage{bm,epsfig,amsthm,url}
\usepackage{indentfirst}
\usepackage{amsfonts}
\usepackage{epstopdf}
\usepackage{cases}
\usepackage[caption=false,font=scriptsize]{subfig}
\usepackage{xcolor}
\usepackage{mathtools}

\makeatletter
\renewcommand{\maketag@@@}[1]{\hbox{\m@th\normalsize\normalfont#1}}%
\makeatother
\usepackage{hyperref}
\usepackage{stfloats}

\newtheoremstyle{mystyle}{}{}{}{}{}{: }{0pt}{\indent \it{\thmname{#1}\thmnumber{ #2}\thmnote{#3}}}
\theoremstyle{mystyle}

\newtheorem{Proposition}{Proposition}

\newtheorem{Remark}{Remark}

\usepackage{amsmath}
\allowdisplaybreaks[4]

\begin{document}

\title{\fontsize{19pt}{26pt}\selectfont Semi-Passive IRS Enabled Sensing with Group Movable Sensors}

\author{{Qiaoyan~Peng,~Qingqing~Wu,~Wen~Chen,~Guangji~Chen,~Ying~Gao,~Lexi Xu,~Shaodan~Ma}
	\thanks{Q. Wu's work was supported by NSFC 62331022, NSFC 62371289, and Shanghai Jiao Tong University 2030 Initiative. The work of G. Chen is supported by the Natural Science Foundation of Jiangsu Province under Grant BK20241455, and the open research fund of National Mobile Communications Research Laboratory, Southeast University (No. 2025D14). The work of W. Chen is supported by Shanghai Kewei 22JC1404000 and 24DP1500500. This work of S. Ma is supported in part by the Science and Technology Development Fund, Macau SAR under Grants 0087/2022/AFJ and 001/2024/SKL, in part by the National Natural Science Foundation of China under Grant 62261160650, and in part by the Research Committee of University of Macau under Grant MYRG-GRG2023-00116-FST-UMDF.
	\textit{(Corresponding author: Qingqing Wu.)}
		
	Q. Peng is with the Department of Electronic Engineering, Shanghai Jiao Tong University, Shanghai 200240, China, and also with the State Key Laboratory of Internet of Things for Smart City, University of Macau, Macao 999078, China (email: qiaoyan.peng@connect.um.edu.mo). Q. Wu, W. Chen, and Y. Gao are with the Department of Electronic Engineering, Shanghai Jiao Tong University, Shanghai 200240, China (e-mail: \{qingqingwu, wenchen, yinggao\}@sjtu.edu.cn).
	G. Chen is with the School of Electronic and Optical Engineering, Nanjing University of Science and Technology, Nanjing 210094, China  and
	also with National Mobile Communications Research Laboratory, Southeast University (email:
	guangjichen@njust.edu.cn).
	L. Xu is with the Research Institute, China United Network Communications Corporation, Beijing 100048, China (e-mail: davidlexi@hotmail.com).
	S. Ma is with the State Key Laboratory of Internet of Things for Smart City, University of Macau, Macao 999078, China (email: shaodanma@um.edu.mo).
	}
}

\maketitle
\begin{abstract}
The performance of the sensing system is limited by the signal attenuation and the number of receiving components. In this letter, we investigate the sensor position selection in a semi-passive intelligent reflecting surface (IRS) enabled non-line-of-sight (NLoS) sensing system. The IRS consists of passive elements and active sensors, where the sensors can receive and process the echo signal for direction-of-arrival (DoA) estimation. Motivated by the movable antenna array and fluid antenna system, we consider the case where the sensors are integrated into a group for movement and derive the corresponding Cramer-Rao bound (CRB). Then, the optimal solution for the positions of the movable sensors (MSs) to the CRB minimization problem is derived in closed form. Moreover, we characterize the relationship between the CRB and system parameters. Theoretical analysis and numerical results are provided to demonstrate the superiority of the proposed MS scheme over the fixed-position (FP) scheme.
\end{abstract}

\section{Introduction}
Traditional mono-static and bi-static base station (BS) sensing systems depend on line-of-sight (LoS) BS-target links \cite{survey}. However, for practical scenarios where targets are in the non-LoS (NLoS) region, sensing performance may degrade significantly, which poses challenges for effective target estimation.

Previous studies have extensively demonstrated the advantages of intelligent reflecting surfaces (IRS) in wireless communication by enabling reconfigurable signal propagation, with recent studies highlighting their potential benefits in sensing \cite{BF, HIRS_EE, FPS}. To further improve sensing performance, a semi-passive IRS has been proposed \cite{IRS_sensing, SPS1, SPS2}, equipped with both passive reflecting elements and active sensors to directly receive and process echo signals propagating through the BS-IRS-target-IRS link. Compared to a fully-passive IRS, the semi-passive IRS is preferable for deployment, as it enables a shorter signal path, reduces signal attenuation, and enhances sensing accuracy. In \cite{SPS1, SPS2}, the impact of system parameters on the Cramer-Rao bound (CRB) was characterized. The results unveiled that CRB decreases with the number of sensors, i.e., increasing the number of sensors helps to improve the point target's direction-of-arrival (DoA) estimation accuracy. To address similar issues caused by large-scale antenna arrays, sparse antenna arrays were proposed to reduce the number of antennas while increasing the antenna spacing \cite{sparse}. However, it may not satisfy different sensing requirements due to the adopted fixed-position (FP) antennas. In contrast, a movable antenna (MA) aided system \cite{MA_sensing} (or fluid antenna system \cite{FAS}) was considered, where the positions of antennas can be optimized, thereby providing new degrees of freedom for improving the sensing performance. The array aperture can be effectively increased by enlarging the antenna movement region with the same number of antennas, which helps achieve a higher angular resolution. Motivated by MA technology \cite{MA-IRS}, we endow the mobility characteristics of MA to the semi-passive IRS, where sensor positions can be optimized, although the optimal positions remain unresolved. Given cost and size constraints in semi-passive IRS-aided sensing systems, it is crucial to optimize sensor positions for maximizing sensing performance without increasing the number of sensors. It can alleviate the requirement for additional sensors, thereby reducing costs and complexity. Moreover, the extent of performance improvement with optimized sensor positions over FP schemes remains unknown, which motivates our work.

In this letter, we focus on parameter estimation in a semi-passive IRS-aided wireless sensing system with optimized sensor positions. Different from the conventional FP scheme, the semi-passive IRS consists of movable sensors (MSs). Instead of moving independently, the sensors are integrated into a group for movement, which significantly reduces the implementation costs and complexity, and is more appealing in practical scenarios \cite{group1, group2}. To this end, we first derive the CRB for DoA estimation and then formulate the corresponding minimization problem by optimizing the sensor position vector (SPV). With the optimal solution, we unveil that traditional DoA estimation algorithms may not be effective for the MS array due to its non-uniformity. Theoretical and numerical results verify our findings and demonstrate the superiority of the MS scheme over the FP scheme in terms of CRB. 

\section{System Model and Problem Formulation}
\begin{figure}[t]
	\centering
	\includegraphics[width=0.62\linewidth]{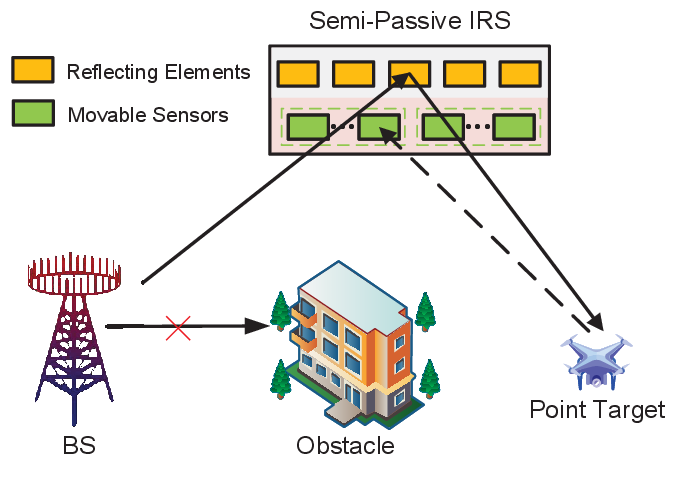}
	\caption{Illustration of the semi-passive IRS enabled sensing system.}
	\label{fig:model}
\end{figure}

We consider a semi-passive IRS-enabled sensing system as illustrated in Fig. \ref{fig:model}, which consists of a BS with $M$ FP antennas, a semi-passive IRS,\footnote{Assuming each IRS is deployed in a separate region without inter-region interference, our results are applicable to a multi-IRS scenario, as in \cite{SPS3}.} and a point-like target. The BS-target direct link is assumed to be blocked due to dense obstacles. The IRS consists of $N$ passive reflecting elements and $K$ MSs, where reflecting elements are used for adjusting the phases of probing signals, while MSs have signal reception and processing capabilities. The total number of sensors is given by $K=LK_l$, where $L \ge 2$ and $K_l \ge 1$ denote the number of groups and the number of sensors within the $l$-th group, respectively. We assume that $M$ and $N$ are even integers, as well as $K$ is an integer, with $M \ge 2$, $N \ge 2$, and $K \ge 2$. The positions of MSs can be adjusted flexibly within the given 1D line segment of length $D$. \footnote{Our results can be extended to the 2D MS array case by alternately optimizing the horizontal and vertical coordinates of the MSs' positions using similar optimization methods in \cite{MA_sensing}.} Let $\mathbf{x} \triangleq [x_{1,1}, \ldots, x_{1,K_1}, \ldots, x_{L,1}, \ldots, x_{L,K_L} ]$ denote the SPV of the MS array, where $x_{l,k} = x_{l,k-1} + d_l$, $1 \le l \le L, 2 \le k \le K_l$ represents the position of the $k$-th sensor within the $l$-th group with the inter-sensor spacing $d_l$. The placement and movement of the MSs will not modify the geometry of the metasurface or affect the positions of the reflecting elements. Moreover, the codebook for reflecting elements can be pre-saved, updating based on changes in the direction of interest.

We assume that all the involved links follow the far-field LoS channel model, which remains static over $T$ snapshots. The steering vector (SV) of the MS array is determined by the SPV $\mathbf{x}$ and the target's DoA $\theta$, which is given by ${\mathbf{b}} ( {{\mathbf{x}, \theta }} )= { [ {{e^{j\frac{{2\pi }}{{{\lambda _{\mathrm{R}}}}}{x_{1,1}}\sin \theta }}, \ldots ,{e^{j\frac{{2\pi }}{{{\lambda _{\mathrm{R}}}}}{x_{L,K_L}}\sin \theta }}}  ]^T}$, where $\lambda _{\mathrm{R}}$ denotes the wavelength. Let $\beta _{\mathrm{IS}} = {\beta _0}{\beta _1} \in \mathbb{C}$ denote complex channel coefficient, where ${\beta _1} = \sqrt{ {{{\lambda_\mathrm{R}^{2}}\kappa }}/({{64{\pi ^3}d_{{\mathrm{IT}}}^4}})}$ represents the signal attenuation due to propagation from the IRS to the target and back to IRS sensors, as well as target scattering, with the IRS-target distance $d_\mathrm{IT}$ and the target's radar cross section $\kappa$. The small-scale fading ${\beta _0} \sim \mathcal{CN} ( {0,1} )$ refers to rapid fluctuation of the received signal. Considering the target is in the far-field region of the semi-passive IRS, the angles-of-departure from the IRS elements to the target are assumed to equal the target's DoA. The target response matrix can be modeled as ${\mathbf{H}} = \beta _{\mathrm{IS}} {\mathbf{b}}( {{\mathbf{x}, \theta}} ){{\mathbf{a}}^T}( {{\theta }} )$, where the SV at the IRS is given by ${\mathbf{a}} ( {{\theta }} ) =[ {{e^{ - \frac{{j\pi ( {N - 1} ){d_{\mathrm{I}}}\sin \theta }}{{{\lambda _{\mathrm{R}}}}}}}, \ldots ,{e^{\frac{{j\pi ( {N - 1} ){d_{\mathrm{I}}}\sin \theta }}{{{\lambda _{\mathrm{R}}}}}}}}  ]^T$ with the inter-element spacing at the semi-passive IRS ${d_\mathrm{I}}$. Since the array centroid is chosen as the reference point, we have
\begin{align}
	\label{symmetry}
	{{{\mathbf{\dot a}}}^H ( {{\theta }} )}{\mathbf{a} ( {{\theta }} )} = {{\mathbf{a}}^H ( {{\theta }} )}{\mathbf{\dot a} ( {{\theta }} )} = 0,
\end{align}
where ${\mathbf{\dot a}}( {{\theta }} ) = \frac{{\partial {\mathbf{a}}}}{{\partial \theta }} = j\pi \frac{{{d_{\mathrm{I}}}}}{{{\lambda _{\mathrm{R}}}}}\cos \theta {{\mathbf{D}}_{\mathrm{a}}}{\mathbf{a}}$ with ${{\mathbf{D}}_{\mathrm{a}}} = \operatorname{diag} ( { - ( {N - 1} ), - ( {N - 3} ), \ldots ,( {N - 1} )} )$.

The channel between BS and reflecting elements at the semi-passive IRS is modeled as ${\mathbf{G}} = {\beta _{{\mathrm{BI}}}}{\mathbf{a}}( {{\theta _{\mathrm{A}}}} ){{\mathbf{c}}^T}( {{\theta _{\mathrm{D}}}} )$, where ${\beta _{{\mathrm{BI}}}} = \sqrt {{{{\lambda_\mathrm{R}^{2}}}}/({{16{\pi ^2}d_{{\mathrm{BI}}}^2}})}$ denotes the distance-dependent path loss with the BS-IRS distance $d_{{\mathrm{BI}}}$. The transmit response vector of the BS and the received response vector of IRS reflecting elements are given by ${\mathbf{c}}( {{\theta _{\mathrm{D}}}} ) = {[ {{e^{ - \frac{{j\pi ( {M - 1} ){d_{\mathrm{B}}}\sin {\theta _{\mathrm{D}}}}}{{{\lambda _{\mathrm{R}}}}}}}, \ldots ,{e^{\frac{{j\pi ( {M - 1} ){d_{\mathrm{B}}}\sin {\theta _{\mathrm{D}}}}}{{{\lambda _{\mathrm{R}}}}}}}} ]^T}$ and ${\mathbf{a}}( {{\theta _{\mathrm{A}}}}) = { [ {{e^{ - \frac{{j\pi ( {N - 1} ){d_{\mathrm{I}}}\sin {\theta _{\mathrm{A}}}}}{{{\lambda _{\mathrm{R}}}}}}}, \ldots ,{e^{\frac{{j\pi ( {N - 1} ){d_{\mathrm{I}}}\sin {\theta _{\mathrm{A}}}}}{{{\lambda _{\mathrm{R}}}}}}}} ]^T}$, respectively, with the angle of departure $\theta _{\mathrm{D}}$, the angle of arrival $\theta _{\mathrm{A}}$, and the inter-antenna spacing at the BS $d_{\mathrm{B}}$.

The signal passing through the BS-sensor link lacks useful information about the target, which can be excluded to improve sensing performance. The corresponding channel information can be obtained offline by the sensors and effectively eliminated before target estimation. Thus, the signal transmitted from the BS towards the IRS, then to the target, and finally reflected to the MSs at $t$-th snapshot, $1 \le t \le T$, is given by ${\mathbf{y}}( {t} ) = {\beta _{{\mathrm{BI}}}}{\beta _{{\mathrm{IS}}}}{\mathbf{b} }(\mathbf{x}, \theta ){{\mathbf{a}}^T}( \theta  ){\mathbf{\Phi }}{\mathbf{a}}( {{\theta _{\mathrm{A}}}}){{\mathbf{c}}^T}( {{\theta _{\mathrm{D}}}} ){\mathbf{w}}s( t ) + {\mathbf{n}}( t )$, where $\mathbf{\Phi} = \operatorname{diag} (\bm{\phi}) = \operatorname{diag} (e^{j \varphi_1},\cdots,e^{j \varphi_N})$ denotes the IRS phase-shift matrix, $s (t)$ represents the unit-power transmitted data, $\mathbf{w}$ denotes the beamformer, and ${\mathbf{n}}( t ) \sim {\mathcal{CN}} ( {{\mathbf{0}},\sigma _{\mathrm{R}}^2{{\mathbf{I}}_{K}}} )$ is the additive white Gaussian noise at the MSs. For DoA estimation, the received signals can be stacked into the following matrix as ${\mathbf{Y}} = [\mathbf{y}(1),\ldots,\mathbf{y}(T)] = {\beta _{{\mathrm{BI}}}}{\beta _{{\mathrm{IS}}}}{\mathbf{B} (\mathbf{x}, \theta )}{\mathbf{S}} + {\mathbf{N}}$, where $\mathbf{S} = [\mathbf{s}(1),\ldots,\mathbf{s}(T)]$ with $\mathbf{s}(t) = \mathbf{w}{s}(t)$, ${\mathbf{B} (\mathbf{x}, \theta )} = {\mathbf{b}}(\mathbf{x}, \theta  ){{\mathbf{a}}^T}( \theta  ){\mathbf{\Phi }}{\mathbf{a}}( {{\theta _{\mathrm{A}}}} ){{\mathbf{c}}^T}( {{\theta _{\mathrm{D}}}} )$, and $\mathbf{N} = [\mathbf{n}(1),\ldots,\mathbf{n}(T)]$. For notational convenience, below we drop $\mathbf{x}$ and $\theta$ in ${\mathbf{B} (\mathbf{x}, \theta )}$. 

Note that CRB is a crucial sensing performance metric for target estimation, which provides a lower bound on the variance of unbiased parameter estimators. According to \cite{SPS2}, the CRB is given by ${\mathrm{CRB}}( \theta ) {\rm =} \frac{{\sigma _{\mathrm{R}}^2}}{{2T{{\left| {{\beta _{{\mathrm{BI}}}}{\beta _{{\mathrm{IS}}}}} \right|}^2} {( {{\operatorname{tr}} ( {{\mathbf{\dot B}}{{\mathbf{R}}}{{{\mathbf{\dot B}}}^H}} ) - \frac{{{{\left| {{\operatorname{tr}} ( {{\mathbf{B}}{{\mathbf{R}}}{{{\mathbf{\dot B}}}^H}} )} \right|}^2}}}{{{\operatorname{tr}} ( {{\mathbf{B}}{{\mathbf{R}}}{{\mathbf{B}}^H}} )}}} )} }}$, where the covariance matrix of the transmit signal is approximated as $\mathbf{R} = \sum\nolimits_{t = 1}^{T} {\mathbf{s}(t)} {\mathbf{s}(t)^{H}} \approx {\mathbf{w}}{{\mathbf{w}}^H}$, and the first-order partial derivative of $\mathbf{B}$ with respect to (w.r.t.) $\theta$ is denoted as $\mathbf{\dot B} = \frac{{\partial \mathbf{B}}}{{\partial \theta }} = ( {\mathbf{\dot b}} (\mathbf{x}, \theta ){{\mathbf{a}}^T} ( \theta ) + {\mathbf{b}} ( \theta ){{{\mathbf{\dot a}}}^T} ( \theta ) ){\mathbf{\Phi }}{\mathbf{a}}( {{\theta _{\mathrm{A}}}} ){{\mathbf{c}}^T}( {{\theta _{\mathrm{D}}}} )$. The partial derivative of $\mathbf{b} (\mathbf{x},\theta)$ w.r.t. $\theta$ is given by ${\mathbf{\dot b} (\mathbf{x},\theta)} = \frac{{\partial {\mathbf{b}}}}{{\partial \theta }} = j\frac{{2\pi }}{{{\lambda _{\mathrm{R}}}}}\cos \theta {{\mathbf{D}}_{\mathrm{b}}}{\mathbf{b}  (\mathbf{x}, \theta )}$ with ${{\mathbf{D}}_{\mathrm{b}}} = \operatorname{diag} ( \mathbf{x} )$. Leveraging the orthogonality property \eqref{symmetry} yields 
\begin{align}
	&{\operatorname{tr}} ({{\mathbf{B}}{{\mathbf{R}}}{{\mathbf{B}}^H}} ) {\rm =}  {\operatorname{tr}} ( {\mathbf{b}} (\mathbf{x}, \theta ){{\mathbf{a}}^T} ( \theta ){\mathbf{\Phi }}{\mathbf{a}}( {{\theta _{\mathrm{A}}}} ){{\mathbf{c}}^T} ( {{\theta _{\mathrm{D}}}} ){\mathbf{w}} {{\mathbf{w}}^H}{\mathbf{c}^*} ( {{\theta _{\mathrm{D}}}} ) \nonumber\\
	&{\rm \times} {{\mathbf{a}}^H} ( {{\theta _{\mathrm{A}}}} ){{\mathbf{\Phi }}^H}{{\mathbf{a}}^*} ( \theta ){{\mathbf{b}}^H} (\mathbf{x}, \theta ) ) \nonumber\\
	&{\rm =} K{ | {{{\mathbf{a}}^T} ( \theta ){\mathbf{\Phi }}{\mathbf{a}} ( {{\theta _{\mathrm{A}}}} ){{\mathbf{c}}^T} ( {{\theta _{\mathrm{D}}}} ){\mathbf{w}}} |^2}, \label{BB}\\ 
	& {\operatorname{tr}} ( {{\mathbf{B}}{{\mathbf{R}}}{{{\mathbf{\dot B}}}^H}} ) {\rm =} {\operatorname{tr}} ( {\mathbf{b}} (\mathbf{x}, \theta ){{\mathbf{a}}^T} ( \theta ){\mathbf{\Phi }}{\mathbf{a}} ( {{\theta _{\mathrm{A}}}} ){{\mathbf{c}}^T} ( {{\theta _{\mathrm{D}}}} ){\mathbf{w}}{{\mathbf{w}}^H}{\mathbf{c}^*}( {{\theta _{\mathrm{D}}}} ) \nonumber\\
	&{\rm \times} {{\mathbf{a}}^H} ( {{\theta _{\mathrm{A}}}} ){{\mathbf{\Phi }}^H} { ( {{{\mathbf{a}}^*} ( \theta ){{{\mathbf{\dot b}}}^H} (\mathbf{x}, \theta ) 
	+ {{{\mathbf{\dot a}}}^*} ( \theta ){{\mathbf{b}}^H} (\mathbf{x}, \theta )} )} ) \nonumber\\
	&{\rm =} - \frac{j{{2\pi }}\cos \theta}{{{\lambda _{\mathrm{R}}}}} \sum _{l = 1}^L \sum _{k = 1}^{K_L} {{x_{l,k}}} {\left| {{{\mathbf{a}}^T}( \theta  ){\mathbf{\Phi }}{\mathbf{a}}( {{\theta _{\mathrm{A}}}} ){{\mathbf{c}}^T}( {{\theta _{\mathrm{D}}}} ){\mathbf{w}}} \right|^2}, \label{BBdot} \\
	& {\operatorname{tr}} ( {{\mathbf{\dot B}}{{\mathbf{R}}}{{{\mathbf{\dot B}}}^H}} ) {\rm =} {\mathbf{\dot b}} (\mathbf{x}, \theta ){{\mathbf{a}}^T} ( \theta ){\mathbf{\Phi }}{\mathbf{a}} ( {{\theta _{\mathrm{A}}}} ){{\mathbf{c}}^T} ( {{\theta _{\mathrm{D}}}} ){\mathbf{w}}{{\mathbf{w}}^H}{\mathbf{c}^*}( {{\theta _{\mathrm{D}}}} ) \nonumber\\
	&{\rm \times} {{\mathbf{a}}^H} ( {{\theta _{\mathrm{A}}}} ){{\mathbf{\Phi }}^H}{{\mathbf{a}}^*} ( \theta ){{{\mathbf{\dot b}}}^H} (\mathbf{x}, \theta ) + {\mathbf{b}} (\mathbf{x}, \theta ){{{\mathbf{\dot a}}}^T} ( \theta ){\mathbf{\Phi }} {\mathbf{a}} ( {{\theta _{\mathrm{A}}}} ){{\mathbf{c}}^T} ( {{\theta _{\mathrm{D}}}} )\nonumber\\
	&{\rm \times} {\mathbf{w}} {{\mathbf{w}}^H}{\mathbf{c}^*} ( {{\theta _{\mathrm{D}}}} )  {{\mathbf{a}}^H} ( {{\theta _{\mathrm{A}}}} ){{\mathbf{\Phi }}^H}{{{\mathbf{\dot a}}}^*} ( \theta ){{\mathbf{b}}^H} (\mathbf{x}, \theta ) \nonumber \\
	&{\rm =} {{4{\pi ^2}}}{\cos ^2}\theta/{{\lambda _{\mathrm{R}}^2}} \sum\nolimits_{l = 1}^L \sum\nolimits_{k = 1}^{K_L} {x_{l,k}^2} { | {{{\mathbf{a}}^T} ( \theta ){\mathbf{\Phi }}{\mathbf{a}} ( {{\theta _{\mathrm{A}}}} ){{\mathbf{c}}^T} ( {{\theta _{\mathrm{D}}}} ){\mathbf{w}}} |^2} \nonumber\\
	&{\rm +} K{| {{{{\mathbf{\dot a}}}^T} ( \theta ){\mathbf{\Phi }}{\mathbf{a}} ( {{\theta _{\mathrm{A}}}} ){{\mathbf{c}}^T} ( {{\theta _{\mathrm{D}}}} ){\mathbf{w}}} |^2}. \label{BdotBdot}
\end{align}

In target tracking scenarios, especially for static or slow-moving targets, the beamforming design for a point target depends on the intended sensing direction. Therefore, we assume that $\theta$ is fixed. Note that enhancing the signal power received at the target is beneficial for improving the sensing accuracy. Under the case where the received signal power is maximized, the probing signal should be directed from the BS towards the IRS and then reflected to the target. Thus, the optimal BS beamformer and IRS phase shift are given by ${\mathbf{w}^\mathrm{opt}} = \sqrt {{{{P_0}}}/{M}} {\mathbf{c}^*}( {{\theta _{\mathrm{D}}}} )$ and ${\bm{\phi}^\mathrm{opt}} = \frac{{{{( {\operatorname{diag}( {{{\mathbf{a}}^T}( \theta  )} ){\mathbf{a}}( {{\theta _{\mathrm{A}}}} )} )}^*}}}{{\left| {\operatorname{diag}( {{{\mathbf{a}}^T}( \theta  )} ){\mathbf{a}}( {{\theta _{\mathrm{A}}}} )} \right|}}$, respectively, where $P_0$ is the transmit power. To facilitate derivation, we use ${\mathrm{CRB}}_{\theta}( \mathbf{x} )$ to represent ${\mathrm{CRB}}( \theta )$, since ${\mathrm{CRB}}( \theta )$ is a function w.r.t. $\mathbf{x}$. With ${\mathbf{w}^\mathrm{opt}}$ and ${\bm{\phi}^\mathrm{opt}}$, we have
\begin{align}
	\label{crb_closed}
	{\mathrm{CRB}}_{\theta}( \mathbf{x} ) = \frac{{\sigma _{\mathrm{R}}^2\lambda _{\mathrm{R}}^2}}{{8{\pi ^2}{{\cos }^2}\theta {P_0}TMK{N^2}{{ | {{\beta _{{\mathrm{BI}}}}{\beta _{{\mathrm{IS}}}}} |}^2}{\operatorname{var}} ( {\mathbf{x}} )}},
\end{align}
where $\operatorname{var} ( {\mathbf{x}} ) \triangleq \frac{1}{K} \sum\nolimits_{l = 1}^L \sum\nolimits_{k = 1}^{K_L} {x_{l,k}^2}  - \frac{1}{{{K^2}}}{( {\sum\nolimits_{l = 1}^L \sum\nolimits_{k = 1}^{K_L} {{x_{l,k}}} } )^2}$ is denoted as the variance function of $\mathbf{x}$. 

Our objective is to minimize the CRB by optimizing the sensor position, which can be formulated as
\begin{subequations}
	\label{prob_CRB}
	\begin{align}
		\mathop {\min }\limits_{\mathbf{x}} \;\;\;\; &{\mathrm{CRB}}_{\theta}( \mathbf{x} ) \\
		\mathrm{s.t.} \;\;\;\; & x_{1,1} \ge 0, x_{L,K_L} \le D, \label{con_x_1}\\
		&x_{l,k} - x_{l,k-1} \ge d_\mathrm{min}, 2 \le k \le K_l, 1 \le l \le L, \label{con_x_2} \\
		&x_{l,1} - x_{l-1, 1} \ge K_{l-1} d_\mathrm{min}, 2 \le l \le L, \label{con_x_3}
	\end{align}
\end{subequations}
where $d_\mathrm{min}$ is the minimum inter-sensor spacing to avoid the coupling effect, which satisfies $0 < d_{{\mathrm{min}}} \le D/(K-1)$. Note that $D \ge (K-1) d_\mathrm{min}$ holds for the feasibility of problem \eqref{prob_CRB}, ensuring the constraints \eqref{con_x_2} and \eqref{con_x_3} are satisfied. Problem \eqref{prob_CRB} is intractable due to its non-convex objective function.
 
\section{Solution and Performance Analysis}
In this section, we focus on the impact of the number and positions of MSs on the CRB and aim to further reduce the CRB by adjusting these positions. Then, we compare the MS scheme with the traditional FP scheme in terms of the CRB. 

It is observed from \eqref{crb_closed} that ${\mathrm{CRB}}_{\theta}( \mathbf{x} )$ decreases with ${\operatorname{var}} ( {\mathbf{x}} )$. To facilitate derivation, we consider the case where each group is equipped with $K_l = \bar{K}$ MSs, denoted by the set $\mathcal{K} \triangleq \{ 1, \cdots , \bar K \}$, and the inter-sensor spacing within each group is set as the minimum spacing, i.e., $d_l = d_\mathrm{min}$, where $1 \le l \le L$. As such, problem \eqref{prob_CRB} is equivalent to
\begin{align}
	\label{prob_var}
	\mathop {\max }\limits_{\mathbf{x}} \;\; \operatorname{var} ( {\mathbf{x}} )
		\;\;\;\;\;\; \mathrm{s.t.} \;\; \eqref{con_x_1}, \eqref{con_x_3}.
\end{align}
Although problem \eqref{prob_var} is still non-convex, we can obtain the optimal solutions in the following proposition. 

\begin{Proposition}
	One optimal solution to problem \eqref{prob_var} is
	\begin{align}
		\label{x_opt_1}
		x_{l,k}^\mathrm{opt} \!\!\!=\!\! \left\{ {\begin{array}{*{20}{l}}
		{\!\!\!\! (\! \bar{K}( {l \!-\! 1} ) \!\!+\! k \!-\! 1){d_{{\mathrm{min}}}},}\\
		{\!\!\!\! D \!\! -\!\!  (\! {K \!\! -\!\!  \bar{K} \! ( l \! - \! 1 \!) \!\! - \! k} \!){d_{{\mathrm{min}}}},}
			\end{array}\begin{array}{*{20}{l}}
				{\!\!\!\!\!\! \forall k \!\! \in \!\! \mathcal{K},\! l \!\! = \!\!  1,\! \ldots ,\! \lfloor \!  {L/2}\!  \rfloor ,}\\
				{\!\!\!\!\!\! \forall k \!\! \in \!\! \mathcal{K},\! l \!\! = \!\!  \lfloor \!  {L/2} \! \rfloor \!\! + \!\! 1,\! \ldots \!,\! L.}
		\end{array}} \right. 
	\end{align}
	When $L$ is odd, another optimal solution to problem \eqref{prob_var} is
	\begin{align}
		\label{x_opt_2}
		x_{l,k}^\mathrm{opt} \!\!\!=\!\! \left\{ {\begin{array}{*{20}{l}}
				{\!\!\!\! (\! \bar{K}( {l \!-\! 1} ) \!\!+\! k \!-\! 1){d_{{\mathrm{min}}}},}\\
				{\!\!\!\! D \!\! -\!\!  (\! {K \!\! -\!\!  \bar{K} \! ( l \! - \! 1 \!) \!\! - \! k} \!){d_{{\mathrm{min}}}},}
			\end{array}\begin{array}{*{20}{l}}
				{\!\!\!\!\!\! \forall k \!\! \in \!\! \mathcal{K},\!  l \!\! = \!\!  1,\! \ldots ,\! \lfloor \!  {L/2} \!  \rfloor \!\! + \!\! 1,}\\
				{\!\!\!\!\!\! \forall k \!\! \in \!\! \mathcal{K},\!  l \!\! = \!\!  \lfloor \!  {L/2} \! \rfloor \!\! + \!\! 2,\! \ldots \!, \! L.}
		\end{array}} \right. 
	\end{align}
\end{Proposition}

{\it{Proof:}}
For initialization, we set ${{\mathbf{x}}^{( 0 )}} = \mathbf{x}$, which satisfies constraints \eqref{con_x_1} and \eqref{con_x_3}. Then, in the $l$-th adjustment, the positions of the $k$-th MSs in $l$-th group is updated by $x_{l + 1,k}^{ ( {l + 1} )} {\rm \leftarrow} {{\mathbf{x}}^\mathrm{opt}} [ {l\bar{K} + k} ], 1 \le l \le \lfloor L/2 \rfloor$, while the others keeps unchanged, i.e., $x_{p,k}^{( {l + 1} )} = x_{p,k}^{( l )},p {\ne} l + 1, k \in \mathcal{K}$. Since $x_{l + 1,k}^{( {l + 1} )} - x_{l,k}^{ ( {l + 1} )} = {{\mathbf{x}}^\mathrm{opt}}[ {l\bar{K} + k} ] - {{\mathbf{x}}^\mathrm{opt}}[ {( {l - 1} )\bar{K} + k} ] = \bar{K}{d_{\mathrm{min} }}$, $x_{l + 2,k}^{(l)} - x_{l + 1,k}^{( l )} \ge \bar{K}D$, and $x_{l + 1,k}^{( l )} - x_{l,k}^{( l )} \ge \bar{K}{d_{\mathrm{min} }}$, we have $x_{l + 2,k}^{( {l + 1} )} - x_{l + 1,k}^{( {l + 1} )} = ( x_{l + 2,k}^{( {l + 1} )} - x_{l,k}^{( {l + 1} )} ) - ( x_{l + 1,k}^{( {l + 1} )} - x_{l,k}^{( {l + 1} )} ) = ( x_{l + 2,k}^{( l )} - x_{l,k}^{( l )} ) - \bar{K}D = ( x_{l + 2,k}^{( l )} - x_{l + 1,k}^{( l )} ) - ( x_{l + 1,k}^{( l )} - x_{l,k}^{( l )} ) - \bar{K}{d_{\mathrm{min} }} \ge \bar{K}{d_{\mathrm{min} }}$. Based on the mathematical induction, it is guaranteed that ${{\mathbf{x}}^{( l )}}$ satisfies the constraints \eqref{con_x_1} and \eqref{con_x_3}. Since $x_{l,1}^{( l )} - x_{l,1}^{( {l - 1} )} \le 0$ and ${x_{l,1}^{( l )} + x_{l,1}^{( {l - 1} )} + ( {\bar{K} - 1} )d - 2\mu ( {{{\bf{x}}^{( l )}},l} )} \le 0$ with $\mu ( {{\bf{x}},p} ) = {1}/{{( {K - \bar K} )}}\sum\nolimits_{l = 1,l \ne p}^L {\sum\nolimits_{k = 1}^{\bar K} {{x_{l,k}}} }$, it follows that $ {\mathop{\rm var}} ( {{{\bf{x}}^{( l )}}} ) - {\mathop{\rm var}} ( {{{\bf{x}}^{( {l - 1} )}}} ) = {{\bar{K}( {K - \bar{K}} )}}/{{{K^2}}}( x_{l,1}^{( l )} - x_{l,1}^{( {l - 1} )} ) ( x_{l,1}^{( l )} +  x_{l,1}^{( {l - 1} )} + ( {\bar{K} - 1} )d - 2\mu ( {{{\bf{x}}^{( l )}},l} ) ) \ge 0$. When $\lfloor L/2 \rfloor +2 \le l \le L$, the sensor position adjustment procedure is similar and omitted for brevity. Note that symmetric transformations preserve the distribution characteristics of data, especially in the derivation of variance. As such, another solution is presented in \eqref{x_opt_2}. 
~$\hfill\blacksquare$

%\begin{figure*}[t]
%	\begin{align}
%		\label{var_x_opt}
%		\operatorname{var}(\mathbf{x}^*) = \left\{ {\begin{array}{*{20}{l}}
%				 \frac{{\bar{K}}}{{12{K^2}}}( {3( {KL - \bar{K}} ){D^2} - 3( {K - 2} )( {KL - \bar{K}} )D{d_\mathrm{min}} + ( {K - 1} )( {L{K^2} - 2LK + 3\bar{K}} )d_\mathrm{min}^2} )\\
%				 {{\frac{1}{{12}}( {3{D^2} - 3( {\bar{K}L - 2} )D{d_{{\mathrm{min}}}} + ( {{\bar{K}^2L^2} - 3\bar{K}L + 2} )d_{{\mathrm{min}}}^2} )},}
%			\end{array}\begin{array}{*{20}{l}}
%				{L \text{ is odd},}\\
%				{L \text{ is even}.}
%		\end{array}} \right. 
%	\end{align}
%	{\noindent} \rule[-0pt]{18cm}{0.05em}
%\end{figure*}

From \eqref{x_opt_1} and \eqref{x_opt_2}, it can be observed that when the number of groups is even, the MSs can be divided into two groups and positioned at both ends of the line segment with an inter-sensor distance $d_{{\mathrm{min}}}$. When the number of groups is odd, the unpaired group should be placed innermost at either end, depending on its initial position before moving.  
\begin{Remark}
	When $D > (K-1)d_\mathrm{min}$, the MS array is a non-uniform array due to the holes between the left and right uniform sub-arrays. Traditional DoA estimation algorithms, e.g., multiple signal classification (MUSIC) \cite{MUSIC}, can be applied to the MS array but with redundant sidelobes. To fully exploit the superiority of the MSs, an augmented virtual array can be constructed and processed via spatial smoothing, compressive sensing, and array interpolation techniques, which further improves the accuracy of DOA estimation \cite{interpolation}.
\end{Remark}

With $x_{l,k}^\mathrm{opt}$, the variance function is given by $\operatorname{var}(\mathbf{x}^\mathrm{opt}) = {{\bar{K}}}/({{12{K^2}}})( 3( {KL - \bar{K}} ){D^2} - 3( {K - 2} )( {KL - \bar{K}} )D{d_\mathrm{min}} + ( {K - 1} )( {L{K^2} - 2LK + 3\bar{K}} )d_\mathrm{min}^2 )$ when $L$ is odd and $( {3{D^2} - 3( {\bar{K}L - 2} )D{d_{{\mathrm{min}}}} + ( {{\bar{K}^2L^2} - 3\bar{K}L + 2} )d_{{\mathrm{min}}}^2} )/12$ when $L$ is even. We define ${g ( {D,K,{d_{{\mathrm{min}}}}, L, \bar{K}} )} \triangleq K\operatorname{var}(\mathbf{x}^\mathrm{opt})$. With $\operatorname{var}(\mathbf{x}^\mathrm{opt})$, the CRB for DoA estimation is
\begin{align}
	\label{crb_ma}
	{\mathrm{CRB}}_{\theta} (\mathbf{x}) \!\!=\!\! \frac{{\sigma _{\mathrm{R}}^2\lambda _{\mathrm{R}}^2}}{{8{\pi ^2}{{\cos }^2} \theta {P_0}TM{N^2}{{| {{\beta _{{\mathrm{BI}}}}{\beta _{{\mathrm{IS}}}}} |}^2} \!{g (\! {D,\! K,\! {d_{{\mathrm{min}}}}},\! L ,\! \bar{K} \!) }}}.
\end{align}

From \eqref{crb_ma}, it can be readily verified that the ${\mathrm{CRB}}_{\theta} (\mathbf{x})$ monotonically decreases with the transmit power $P_0$, the number of transmit antennas $M$, the number of IRS passive elements $N$, and $D$ for $D \ge (K-1) {d_{{\mathrm{min}}}}$, whereas it monotonically increases with $d_{{\mathrm{min}}}$ for $0 < d_{{\mathrm{min}}} \le D/(K-1)$. In the following proposition, we characterize the impact of the number of groups on the CRB.
\begin{Proposition}
	\label{crb_K}
	${\mathrm{CRB}}_{\theta}( \mathbf{x} )$ monotonically decreases with $L$ except when $\bar{K} = 1$ and $D$ is an even multiple of $d_\mathrm{min}$.
\end{Proposition}
{\it{Proof:}}
When $L$ is odd, we define ${g_{\mathrm{o}}}( L ) \triangleq {{\bar{K}}}/({{12{K}}}) ( 3( {KL - \bar{K}} ){D^2} - 3( {K - 2} )( KL - \bar{K} )D{d_\mathrm{min}} + ( K - 1 )( L{K^2} - 2LK + 3\bar{K} )d_\mathrm{min}^2 )$. When $L$ is even, we define ${g_{\mathrm{e}}}( L ) \triangleq {K}/{12} ( 3{D^2} - 3( {\bar{K}L - 2} )D{d_{{\mathrm{min}}}} + ( {\bar{K}^2L^2} - 3\bar{K}L + 2)d_{{\mathrm{min}}}^2 )$. Note that ${\mathrm{CRB}}_{\theta}( \mathbf{x} )$ is inversely proportional to ${g_{\mathrm{o}}}( L )$ or ${g_{\mathrm{e}}}( L )$, respectively. Since ${g_{\mathrm{o}}}( L ) - {g_{\mathrm{o}}}( {L - 2} ) > 0$ for $K \ge 5$ and ${g_{\mathrm{e}}}( L ) - {g_{\mathrm{e}}}( {L - 2} ) > 0$ for $K \ge 4$. Thus, ${\mathrm{CRB}}_{\theta}( \mathbf{x} )$ monotonically decreases with $L$ when $L$ is even or odd. When $L$ is odd, we have ${g_{\mathrm{o}}}( L ) - {g_{\mathrm{e}}}( {L - 1} ) = {\bar{K}{f_1}( D )/( {12L} )} \ge 0$, where ${f_1}( D ) = 3( {L - 1} ){D^2} - 6( {L - 1} )( {K - 1} ){d_\mathrm{min}}D + ( 3( {L - 1} ){K^2} + ( {\bar{K} - 6L + 6} )K + 2L - 3 )d_\mathrm{min}^2$. It follows that ${g_{\mathrm{o}}}( L ) = {g_{\mathrm{e}}}( {L - 1})$ if only if $D = (L - 1) {d_{\mathrm{min}}} $.
~$\hfill\blacksquare$

It is observed from Proposition \ref{crb_K} and \eqref{crb_ma} that more passive elements and sensors should be deployed at the semi-passive IRS for CRB minimization. In the following, we further characterize the relationship among $N$, $L$, and $\bar{K}$.

\begin{Remark}
	Assuming constraints on power/cost/total number for passive elements and movable sensors, i.e., ${W_1}N + {W_2}L = Q$, we define $p(\tilde{L}) \triangleq N^2 {g ( {D,K,{d_{{\mathrm{min}}}}, \tilde{L}, \bar{K}} )}$, where $\tilde L$ is the continuous value of $L$, $Q$ is the total budget, $W_1$ and $W_2$ denote the weight of each passive element and movable sensor group, respectively. Note that maximizing $p(L)$ is equivalent to minimizing ${\mathrm{CRB}}_{\theta}( \mathbf{x} )$. By setting the first-order partial derivative of $p(\tilde{L})$ w.r.t. ${\tilde L}$ to zero, we can obtain the solutions from $3d_{\min }^2{{\bar K}^2}{{\tilde L}^4} - 6\bar K{d_{\min }} ( D + {d_{\min }} ){{\tilde L}^3} +  ( 2d_{\min }^2 + 6D{d_{\min }} + 3{D^2}  ){{\tilde L}^2} + 3d_{\min }^2 + 6D{d_{\min }} + 3{D^2} =0$ when $L$ is odd, and from $- 5{W_2}\bar Kd_{\min }^2{{\tilde L}^3} + 3\bar K{d_{\min }} ( Q{d_{\min }} + 4{W_2}{d_{\min }} + 4D{W_2} ) {{\tilde L}^2} - 3 ( 2\bar K{d_{\min }} ( {d_{\min }} + D ) Q + {W_2} ( 2d_{\min }^2 + 6D{d_{\min }} + 3{D^2} ) ){\tilde L} + Q ( 2d_{\min }^2 + 6D{d_{\min }} + 3{D^2}  ) = 0$ when $L$ is even. Considering the rounding technique, the optimal $L$ can be obtained from these solutions and $(D+{d_{\min }})/(\bar{K}{d_{\min }})$ or $2$, by comparing the corresponding values of the objective function.
\end{Remark}

To demonstrate the superiority of the MS scheme, we compare it with the conventional FP scheme. For FP scheme with half-wavelength inter-sensor spacing, the position of the $k$-th sensor is given by ${x_{\mathrm{FP},k}} = {{( {k - 1} ){\lambda _{\mathrm{R}}}}}/{2}$. Then, the variance function can be expressed as $\operatorname{var}(\mathbf{x_{\mathrm{FP}}}) = {{\lambda _{\mathrm{R}}^2( {K + 1} )( {K - 1} )}}/{{48}}$ and the resultant CRB is given by ${\mathrm{CRB}_\mathrm{FP}} ( \theta ) = \frac{{6\sigma _{\mathrm{R}}^2}}{{{\pi ^2}{{\cos }^2}\theta {P_0}TM{N^2}{{\left| {{\beta _{{\mathrm{BI}}}}{\beta _{{\mathrm{IS}}}}} \right|}^2} ( {K^3 - K} )}}$. To facilitate comparison, let $N_1 (N_2)$ and $K_1 (K_2)$ denote the number of IRS reflecting elements and the number of sensors under the FP (MS) scheme, respectively. The CRB reduction ratio is defined as $f(L) \triangleq ({{\mathrm{CRB}_\mathrm{FP}} - {\mathrm{CRB}}_{\theta}( \mathbf{x} )})/{{\mathrm{CRB}_\mathrm{FP}}}$. In the following proposition, we compare the two schemes in terms of CRB to demonstrate the effectiveness of the MS scheme.
\begin{Proposition}
	\label{pro_CRB}
	When $N_1 = N_2 = N$ and $K_1 = K_2 = K = \bar{K}L$, the CRB reduction ratio monotonically increases as $L$ decreases, which is upper-bounded by $f(L) \le  1 - \frac{{3\lambda _{\rm{R}}^2 ( {3{{\bar{K}}^2} - 1} )}}{{32{D^2} - 32 ( {3\bar{K} - 2} )D{d_\mathrm{min}} + 4 ( {3\bar{K} - 1} ) ( {9\bar{K} - 5} )d_\mathrm{min}^2}}$ when $L$ is odd, and $f(L) \le 1 - \frac{{\lambda _{\rm{R}}^2 ( {4{{\bar{K}}^2} - 1} )}}{{12{D^2} - 8 ( {\bar{K} - 1} ) ( {3D + {d_\mathrm{min}}} ){d_\mathrm{min}}}}$ when $L$ is even.
%	\begin{align}
%		f(L)  \le  \left\{ {\begin{array}{*{20}{l}}
%				{{ 1 - \frac{{3\lambda _{\rm{R}}^2 ( {3{{\bar{K}}^2} - 1} )}}{{32{D^2} - 32 ( {3\bar{K} - 2} )D{d_\mathrm{min}} + 4 ( {3\bar{K} - 1} ) ( {9\bar{K} - 5} )d_\mathrm{min}^2}}},}\\
%				{{ 1 - \frac{{\lambda _{\rm{R}}^2 ( {4{{\bar{K}}^2} - 1} )}}{{12{D^2} - 8 ( {\bar{K} - 1} ) ( {3D + {d_\mathrm{min}}} ){d_\mathrm{min}}}}},}
%			\end{array}\begin{array}{*{20}{l}}
%				{ L \text{ is odd},}\\
%				{ L \text{ is even}.}
%		\end{array}} \right.
%	\end{align}
\end{Proposition}

{\it{Proof:}}
When $L$ is odd, $f(L)$ is represented as $f_\mathrm{o} (L)$. Since $f_\mathrm{o} (L)$ monotonically decreases with $K$, we have $f_\mathrm{o} (L)$ is maximized at $L = 3$. When $L$ is even, we use $f_\mathrm{e} (L)$ to represent $f(L)$. Since $f_\mathrm{e} (L)$ monotonically decreases with $L$, it follows that $f_\mathrm{e} (L)$ is maximized at $L = 2$. 
~$\hfill\blacksquare$

It is readily verified from Proposition \ref{pro_CRB} that the upper bound of the CRB gap between MS and FP schemes increases as $D$ increases and as $d_\mathrm{min}$ decreases, except when $\bar{K} = 1$ and $L$ is even, where it is independent of $d_\mathrm{min}$. The results suggest that the practical operating region for the MS scheme with grouping can be extended by increasing the length of the line segment or decreasing the inter-sensor spacing, since it helps increase the movement range, spatial resolution, and flexibility of the array, thereby improving the sensing performance. Furthermore, Proposition \ref{pro_CRB} demonstrates that the MS scheme is more appealing in practical systems, especially when $K$ is small. This is attributed to the high flexibility of the MS scheme, which allows it to fully exploit the impact of positions to enhance sensing performance. 

Based on Proposition \ref{crb_K} and \ref{pro_CRB}, it is observed that to achieve the same CRB with the same number of sensors, i.e., $K_1 = K_2$, the required number of IRS reflecting elements $N_1$ is larger than $N_2$ when $D > (K-1) d_\mathrm{min}$. Similarly, to achieve the same CRB with the same number of IRS reflecting elements, i.e., $N_1 = N_2$, the required number of sensors $K_1$ is larger than $K_2$. These results highlight the potential of the MS scheme as a cost-efficient solution for CRB minimization. 

\section{Simulation Results}
\begin{figure}[t]
	\centering
	\includegraphics[width=0.57\linewidth]{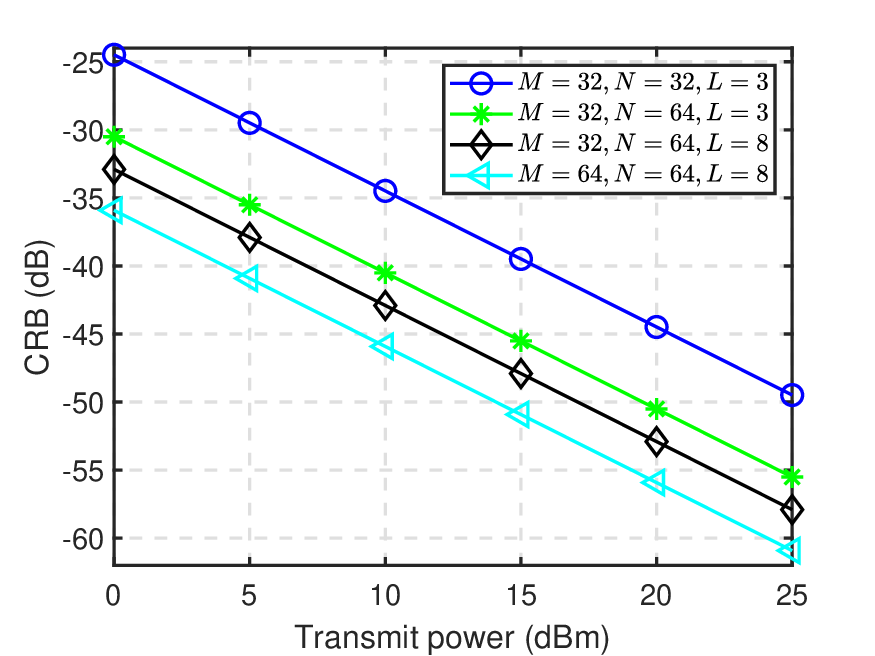}
	\caption{CRB versus transmit power $P_0$.}
	\label{fig:CRB_P}
\end{figure}
\begin{figure}[t]
	\centering
	\includegraphics[width=0.57\linewidth]{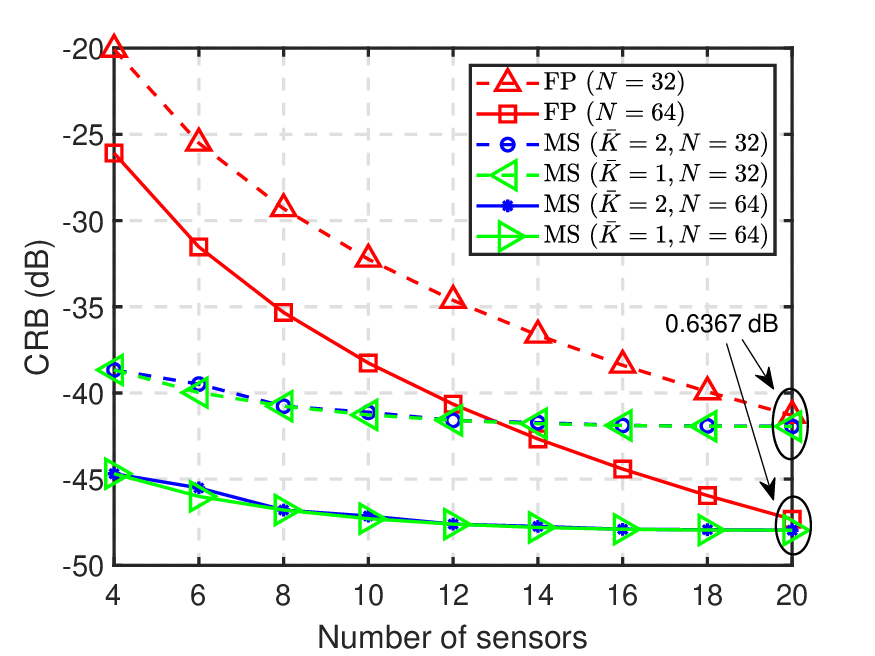}
	\caption{CRB versus number of sensors $K$.}
	\label{fig:CRB_K}
\end{figure}
\begin{figure}[t]
	\centering
	\includegraphics[width=0.57\linewidth]{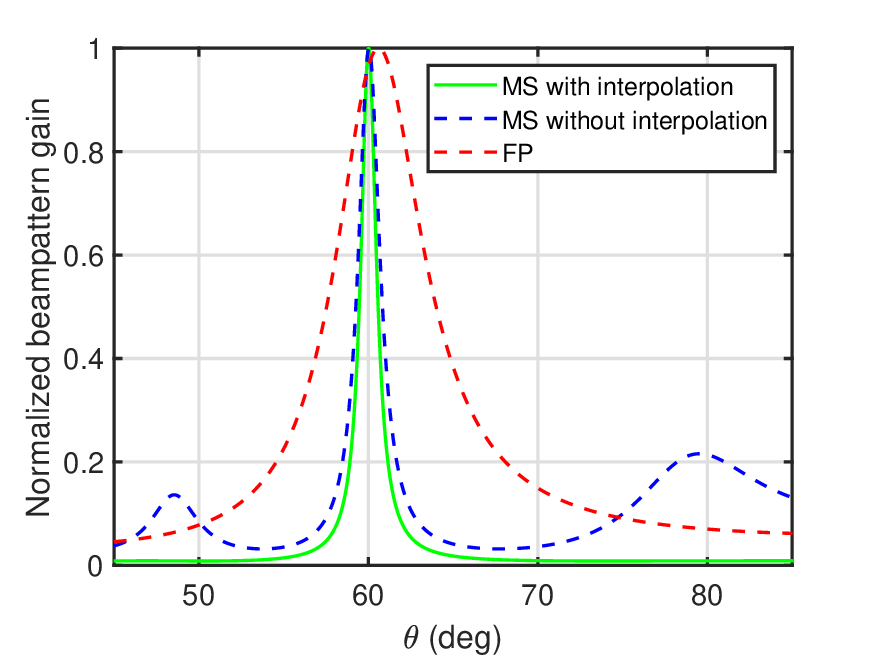}
	\caption{Beampattern versus target's DoA $\theta$.}
	\label{fig:beampattern}
\end{figure}

This section presents numerical results for the DoA estimation of the considered system. The BS-IRS distance and the IRS-target distance are set as $d_\mathrm{BI} = 60$ meter (m) and $d_\mathrm{IT} = 20$ m, respectively. Unless otherwise specified, other system parameters are set as follows: $\kappa = 7$ dBsm, $\theta = 60^\circ$, $T = 64$, $\lambda_{\mathrm{R}} = 0.2$ m, ${d}_\mathrm{min} = \lambda_{\mathrm{R}}/2$, $D = 10 \lambda_{\mathrm{R}}$, and $\sigma_{\mathrm{R}}^2 = -90$ dBm. The CRB, originally in $\deg ^2$, is converted to dB for better visualization.

To evaluate the performance of the MS scheme, we plot the CRB versus the transmit power under different system setups when $\bar{K} = 2$ in Fig. \eqref{fig:CRB_P}. First, it is observed that the CRB decreases linearly with $P_0$ since it is inversely proportional to $P_0$ in \eqref{crb_ma}. Second, The CRB decreases as $M$ increases due to higher transmit beamforming gain and enhanced spatial diversity. Third, a lower CRB is achieved with a larger number of IRS reflecting elements $N$ thanks to higher passive beamforming gain, which improves the received signal power and thus enhances sensing performance. Moreover, we observe that the CRB decreases with $K$. This is expected because an increased number of MSs allows for higher array gain and better resolution of the signal's spatial characteristics, thereby leading to more precise DoA estimation. 

In Fig. \ref{fig:CRB_K}, we study the impact of the number and positions of sensors by plotting the CRB versus $K$ when $M=32$ and $P_0=15$ dBm. To facilitate comparison, we consider the conventional ULA with the FP scheme at half-wavelength inter-sensor spacing. One can observe that the MS scheme outperforms the FP scheme for both $N = 32$ and $N = 64$ with a performance gap of 0.6367 dB when $K = 20$, which is more pronounced for smaller $K$. When $K$ is small, each additional MS significantly increases the variance due to the large positional differences among MSs. The variance keeps increasing with $K$, however, its growth rate decreases since the positional differences between newly added and existing MSs decrease, resulting in a more uniform spatial distribution. Moreover, we observe that the MS scheme with $\bar{K} = 2$ achieves CRB performance comparable to the case without grouping. Although increasing the number of groups may result in a slight performance loss, it offers substantial benefits in terms of reduced optimization complexity and simplified movement control.

In Fig. \ref{fig:beampattern}, we plot the beampatterns versus DoA $\theta$ when $K=8$, $\bar{K} = 2$, $M=N=32$, and $P_0=15$ dBm. We adopt the FP scheme as in Fig. \ref{fig:CRB_K}. The beampatterns under all the schemes are estimated by the MUSIC method \cite{MUSIC}. To further improve the estimation accuracy, we apply the array interpolation method to transform the MS array into a uniform virtual array. According to \cite{centro-symmetric}, we fill the holes where the sensors are missing and construct an array with a centro-symmetric geometry. We observe that the MS schemes with and without array interpolation not only focus their mainlobes more accurately towards the target ($\theta = 60^\circ$) but also produce a narrower main beam than the FP scheme, which demonstrates superior performance enabled by optimized sensors' positions. Moreover, the beampattern under the MS scheme using the array interpolation-based MUSIC method has a narrower main beamwidth and suppressed sidelobes, resulting from the increased aperture of the virtual array. Consequently, it enables more precise energy focusing in the desired direction, reduces signal spread, and thereby enhances spatial resolution and DoA estimation accuracy.

\section{Conclusion}
\label{Conclusion}
In this letter, we investigated CRB minimization in a sensing system aided by semi-passive IRS with group MSs at optimized positions. The CRB for DoA estimation was obtained as an inverse function of the variance of MS positions. With the closed-form expression of the optimal sensor positions, we unveiled that traditional DoA estimation algorithms can be applied but are ineffective for the non-uniform MS array. Moreover, we derived an upper bound on the maximum CRB gain of the MS scheme over the FP scheme. Simulation results validated the theoretical findings and demonstrated the effectiveness of the semi-passive IRS with the optimized sensor positions for improving the sensing performance.

\bibliographystyle{IEEEtran}
\bibliography{refs.bib} 

@article{FPS,
	author       = {Xianxin Song and others},
	title        = {Intelligent Reflecting Surface Enabled Sensing: {C}ram{\'{e}}r-{R}ao
	Bound Optimization},
	journal      = {{IEEE} Trans. Signal Process.},
	volume       = {71},
	pages        = {2011--2026},
	year         = {2023},
	month        = {May}
}

@article{MA_sensing,
	author       = {Wenyan Ma and others},
	title        = {Movable Antenna Enhanced Wireless Sensing via Antenna Position Optimization},
	journal      = {{IEEE} Trans. Wireless Commun.},
	volume       = {23},
	number       = {11},
	pages        = {16575--16589},
	year         = {2024},
	month		 = {Nov.}
}

@article{MUSIC,
	title={Multiple emitter location and signal parameter estimation},
	author={Schmidt, Ralph},
	journal={{IEEE} Trans. Antennas Propag.},
	volume={34},
	number={3},
	pages={276--280},
	year={1986},
	month={Mar.}
}

@article{interpolation,
	author       = {Chengwei Zhou and others},
	title        = {Direction-of-Arrival Estimation for Coprime Array via Virtual Array
	Interpolation},
	journal      = {{IEEE} Trans. Signal Process.},
	volume       = {66},
	number       = {22},
	pages        = {5956--5971},
	year         = {2018},
	month	     = {Nov.}
}

@article{SPS1,
	author       = {Meng Hua and
	Qingqing Wu and
	Wen Chen and
	Zesong Fei and
	Hing Cheung So and
	Chau Yuen},
	title        = {Intelligent Reflecting Surface-Assisted Localization: {Performance}
	Analysis and Algorithm Design},
	journal      = {{IEEE} Wireless Commun. Lett.},
	volume       = {13},
	number       = {1},
	pages        = {84--88},
	year         = {2024},
	month		 = {Jan.}
}

@article{SPS2,
	author       = {Qiaoyan Peng and
	Qingqing Wu and
	Wen Chen and
	Shaodan Ma and
	Ming{-}Min Zhao and
	Octavia A. Dobre},
	title        = {Semi-Passive Intelligent Reflecting Surface-Enabled Sensing Systems},
	journal      = {{IEEE} Trans. Commun.},
	volume       = {72},
	number       = {12},
	pages        = {7674--7688},
	year         = {2024},
	month        = {Dec.}
}

@article{centro-symmetric,
	author       = {Carine El Kassis and others},
	title        = {Advantages of nonuniform arrays using root-{MUSIC}},
	journal      = {Signal Process.},
	volume       = {90},
	number       = {2},
	pages        = {689--695},
	year         = {2010},
	month		 = {Feb.}
}

@article{FAS,
	title={Fluid antenna systems},
	author={Wong, Kai-Kit and others},
	journal={IEEE Trans. Wireless Commun.},
	volume={20},
	number={3},
	pages={1950--1962},
	year={2021},
	month={Mar.}
}

@article{survey,
	author       = {Qingqing Wu and others},
	title        = {Intelligent Surfaces Empowered Wireless Network: {Recent} Advances and
	the Road to {6G}},
	journal      = {Proc. {IEEE}},
	volume       = {112},
	number       = {7},
	pages        = {724--763},
	year         = {2024},
	month		 = {Jul.}
}

@article{HIRS_EE,
	author       = {Qiaoyan Peng and
	Qingqing Wu and
	Guangji Chen and
	Ruiqi Liu and
	Shaodan Ma and
	Wen Chen},
	title        = {Hybrid Active-Passive {IRS} Assisted Energy-Efficient Wireless Communication},
	journal      = {{IEEE} Commun. Lett.},
	volume       = {27},
	number       = {8},
	pages        = {2202--2206},
	year         = {2023},
	month		 = {Jul.}
}

@article{sparse,
	title={Sparse antenna array design for {MIMO} active sensing applications},
	author={Roberts, William and others},
	journal={IEEE Trans. Antennas Propagat.},
	volume={59},
	number={3},
	pages={846--858},
	year={2011},
	month={Mar.}
}

@article{BF,
	author       = {Guangji Chen and
	Qingqing Wu and
	Ruiqi Liu and
	Jingxian Wu and
	Chao Fang},
	title        = {{IRS} Aided {MEC} Systems With Binary Offloading: {A} Unified Framework
	for Dynamic {IRS} Beamforming},
	journal      = {{IEEE} J. Sel. Areas Commun.},
	volume       = {41},
	number       = {2},
	pages        = {349--365},
	year         = {2023},
	month		 = {Feb.}
}

@article{IRS_sensing,
	author       = {Xiaodan Shao and others},
	title        = {Intelligent Reflecting Surface Aided Wireless Sensing: {Applications}
	and Design Issues},
	journal      = {{IEEE} Wireless Commun.},
	volume       = {31},
	number       = {3},
	pages        = {383--389},
	year         = {2024},
	month		 = {Jun.}
}

@article{group1,
	title={Movable antenna-aided hybrid beamforming for multi-user communications},
	author={Zhang, Yichi and others},
	journal={IEEE Trans. Veh. Technol.},
	year={2025, Early Access},
	note={doi: 10.1109/TVT.2025.3533078},
	doi={10.1109/TVT.2025.3533078}
}

@article{group2,
	title={Group Movable Antenna With Flexible Sparsity: {Joint} Array Position and Sparsity Optimization},
	author={Lu, Haiquan and others},
	journal={IEEE Wireless Commun. Lett.},
	volume={13},
	number={12},
	pages={3573--3577},
	year={2024},
	month={Dec.}
}

@article{SPS3,
	author       = {Yuan Fang and others},
	title        = {Multi-{IRS}-Enabled Integrated Sensing and Communications},
	journal      = {{IEEE} Trans. Commun.},
	volume       = {72},
	number       = {9},
	pages        = {5853--5867},
	year         = {2024},
	month		 = {Sep.}
}

@article{MA-IRS,
	title={Exploiting Movable-Element {STARS} for Wireless Communications},
	author={Zhao, Jingjing and others},
	journal={arXiv preprint arXiv:2412.19974},
	year={2024}
}

\end{document}